# Control of poultry chicken malaria by surface functionalized amorphous nanosilica

Dipankar Seth[1,2], Nitai Debnath[2], Ayesha Rahman[2,3], Sunit Mukhopadhyaya[1], Inga Mewis[3], Christian Ulrichs[3], R. L. Bramhachary[4] and Arunava Goswami[2]

[1]Department of Veterinary Pathology, West Bengal University of Animal and Fisheries Sciences, 37 Kshudiram Bose Sarani, Kolkata 700 037, West Bengal, India.

[2]Biological Sciences Division, Indian Statistical Institute, 203 B.T. Road, Kolkata- 700 108, West Bengal, India.

[3]Humboldt-Universität zu Berlin, Institut für Gartenbauwissenschaften, Fachgebiet Urbaner Gartenbau, Lentzeallee 55, 14195 Berlin, Germany.

[4]21B Motijheel, Kolkata, West Bengal, India 700 074.

Correspondence should be addressed to Arunava Goswami agoswami@isical.ac.in

**Surface modified amorphous nanoporous silica molecules with hydrophobic as well as hydrophilic character can be effectively used as therapeutic drug for combating chicken malaria in poultry industry. The amorphous nanosilica was developed by top-down approach using volcanic soil derived silica as source material. Amorphous silica has long been used as feed additive for poultry industry and considered to be safe for human consumption by WHO and USDA. The basic mechanism of action of these nanosilica molecules is mediated by the physical absorption of VLDL, serum triglycerides and other serum cholesterol components in the lipophilic nanopores of nanosilica. This reduces the supply of the host derived cholesterol, thus limiting the growth of the malarial parasite *in vivo*.**

Avian malaria has a worldwide distribution and is of great economic significance for the poultry industry. Organisms such as *P. gallinaceum, P. juxtanucleare and P. durae* may cause up to 90% mortality in poultry. Incidentally, birds with avian malaria have also been used as model systems for studying the pathogenesis and treatment of malaria in humans[1]. Coma and death may occur quickly when the parasite burden is high[2]. Experimental infection in Van Cobb breed of the poultry chicken results in peak parasitemia five to six days post-infection in 100% birds. Biochemical profiles revealed increased serum activities of different classes of cholesterol. Due to strain differences in susceptibility, different anti-malarial drugs are usually prescribed[3]. The major problems in such medication schedule are cumbersome dosage, field applicability and cost-effectiveness. The consumers often turn refractory to the use of these toxic drug applications and the possibility of residual toxicity often affects the pricing of the broilers. Therefore, there is an urgent need for development of an eco-friendly and better therapeutic intervention mechanism for combating malaria in broiler chickens.

Apicomplexan parasites *Plasmodium falciparum*, *P. gallinaceum* and *Toxoplasma gondii* employ similar strategies for development inside the blood cells of the vertebrate host. Parasite-induced synthesis of lipids and depletion of host cholesterol is critical for intra-erythrocytic proliferation of *P. falciparum* and *Toxoplasma* gondii. They utilize to their advantage the increased host lipid[4-10]. In this paper, we demonstrate that surface modified hydrophobic as well as lipophilic nanosilica could be effectively used as novel drugs for treatment of chicken malaria. Amorphous silica used in this study is considered to be safe for human use by WHO and USDA. It has long been used in poultry industry as feed additives. Amorphous silica, being hygroscopic and non-lipophilic in nature, could not perform the aforementioned additional role as a drug. In order to impart the hydrophobicity and lipophilicity as surface characters and also to lower the dose of the silica as a proposed drug (by increasing the surface to volume ratio), amorphous silica from natural sources like volcanic soil were broken into the nanometer range and then surface functionalized.

30 day old Van Cobb breed chicken were artificially infected with *P. gallinaceum* and after seven days they were sacrificed for studying gross pathological changes and are reported below. Figure 1 (panel I) shows the enlargement and pigmentation changes in the artificially infected chicken [day 7 after infection; sub-panel i(2)] liver vis a vis control [day 7; sub-panel i(1)]. Whereas, the panel I sub-panel ii(2) shows the gross morphology of the spleen from the artificially infected chickens (on day 7) vis a vis the spleen of the control chickens [panel I, sub-panels: ii(2)]. The study was conducted also on other organs like kidney, heart, brain (data not shown). These data show that the artificial infection process produced typical changes observed along with the malaria infection in chickens.

The number of the different intra-erythrocytic stages of the parasites in chicken blood before and after introduction of the nanosilica was counted. Figure 1 Panel II and III show the changes in parasite levels in the infected chickens before and after the treatment of nanosilica, as determined by the blood smear analysis. The data show that with the increase in the parasitemia level, both the intra-erythrocytic stages (trophozoites and schizonts) increases and following simultaneous increase typical to the malarial parasite development inside the chicken RBC (data from control chicken not shown). But after the administration of the nanosilica (marked by red arrows), the number of schizonts go down and thereafter the number of trophozoites becomes lesser than in the control (data not shown) and as a result total parasite burden is reduced significantly (panel II). This shows that due to the presence of the nanosilica in the chicken body, the intra-erythrocytic stages do not develop properly. Fig 1, panel III, show the researchers administering malarial parasite via i. p. (sub-panel A) and i. v. (sub-panel B) route while sub-panel C shows the normal chicken (standing) and malaria infected chicken after 5 days of artificial infection showing signs of posterior paralysis.

Figure 2 shows the changes in the level of different kinds of serum cholesterols as well as the total cholesterol content in the control (blue bars) vis a vis malaria infected (pink bars) and infected chickens treated with 0.025 mg / ml nanosilica twice a day (orange bar). The following table shows the values of the different serum cholesterol components.

Table 1. The different serum cholesterol components as well as the total content were measured. In the malaria infected chicken serum, significant increase in the amount of the serum cholesterol were observed. On application of nanosilica (0.025 mg/ ml) cholesterol levels were found to be reduced back to normal levels.

| Cholesterol Types (7 days post infection) | Control Chicken Serum (mgm / dl) | Malaria Infected Chicken serum (mgm / dl) | Nanosilica treated Malaria Infected Chicken serum (mgm / dl) |
|---|---|---|---|
| Serum Cholesterol | 130 ± 4.46 | 194 ± 3.12 | 151 ± 1.21 |
| Serum HDL Cholesterol | 78 ± 3.01 | 68 ± 1.1 | 82 ± 1.86 |
| Serum LDL Cholesterol | 43 ± 2.93 | 86 ± 2.03 | 59 ± 2.08 |
| Serum VLDL Cholesterol | 9 ± 1.12 | 40 ± 2.13 | 10 ± 1.72 |
| Serum Triglycerides | 31 ± 2.47 | 225 ± 3.07 | 39 ± 5.47 |

The results show that during the course of the malaria infection process all the serum components increase. VLDL and serum triglycerides are most notable amongst them. But after the administration of the lipophilic nanosilica at the particular dose mentioned, the levels of these two serum lipids as well as other lipid constituents come back to the normal physiological level. This is essential for the quick recovery of the chickens to normal health. The application of the higher doses of the nanosilica depletes the cholesterol level completely and impedes the process of recovery of the chickens (data not shown). As cholesterol is necessary for a number of normal physiological processes, maintenance of the requisite physiological level has to be taken into consideration while designing the drug dosage.

Figure 3 shows the measurement of changes in the different isoenzymes after the administration of the nanosilica in malaria infected chickens and we studied the changes in different organs. In all the panels, A represents the level of the isoenzymes in normal chicken, B represents the level in the malaria infected chickens while panel C shows the level of the same isoenzymes in the infected chickens treated with nanosilica at the physiological dose mentioned earlier. It is evident from the data that there are non-specific changes in the levels of different enzymes tested. This indicates that there will be no severe side effects due to the application of the nanosilica as drug to the poultry chickens. It could be concluded from these sets of results that malaria infection and subsequent treatment by nanosilica brings about organ specific changes in different organs of the chickens. We found that 80 ± 3% of the infected chicken (n=15) treated with nanosilica recovered from the disease while 100% of the untreated infected chickens (n=15) died within 7-10 days of inoculation. The parasite burden was reduced by margin of 86±2 % compared to untreated infected birds (n=15).

Drug pool for combating malarial parasite is decreasing day by day due to various genetic (viz. generation of drug resistant malarial parasite etc.) and environmental factors (e.g., failure of the mosquito eradication programs). Broiler chicken industries are often worst hit by the surge of the chicken malaria which causes huge morbidity (affecting the growth and thus uniformity of the flock, which finally affects the product pricing and market potential) as well as mortality especially in young chickens. Once the poultry industry is hit by such a disease, the growers have to use certain standard drugs but consumers face the possible side effects of the residual drugs. Therefore a new problem arises and many good poultry farms decide the remove the whole stock for fear of losing reputation of their brand name and product in the market. In order to avoid this situation, drugs against conserved malarial proteins like P0 and SEC65 homologue of *P. falciparum*[11-16] were tried. However, a danger of developing auto-immune disease due to the conservation of the protein residues between host and parasite proteins will always remain present in this approach. We therefore looked for a drug source or lead molecule which has already been used by the poultry industry as feed for a long time and is also considered inert in nature. Naturally occurring amorphous silica is used by poultry industries for a long time and is considered to be safe for human consumption by different regulatory agencies worldwide. But the major problem with amorphous silica is its hygroscopic nature; this does not allow it to absorb other substances once it absorbs water. In order to circumvent this problem, we decided to first increase the surface area by breaking them to the nano-meter range and then modify the surface properties described elsewhere in detail[17-19]. As a result, the nanosilica became hydrophobic as well as lipophilic in nature. These nanosilica possess nanopores and due to their lipophilic nature they could absorb lipids non-specifically via physio-sorption. These particles have been used in the present set of experiments as drugs to mop up the excess amount of the host serum cholesterol lipids which is used by the malarial parasite mainly for their intra-erythrocytic growth. The results show clearly that these nanosilicas at the doses mentioned earlier in the text could be used as excellent therapeutic agent against chicken malaria and will be a very valuable tool for the broiler industry worldwide. To the best of our knowledge, this report stands as a precursor in this area of research worldwide. Furthermore, our preliminary findings (A. Rahman *et al*., unpublished data) also suggest two nanoparticles (AL60102 and AL60106; derived from higher plant stem derived nanosilica) are capable of reducing certain classes of lipo-proteins (in native form) present in the silkworm larval hemolymph. Here, too, infection by nuclear polyhedrosis virus (*BmNPV*), a scourge in silkworm industry enhances the level of certain lipids, part of which is reduced by nanosilica. We are now trying to determine the level of nanosilica to be introduced to silkworm larval body to maintain the physiological level of these lipoproteins at the physiological level. The situation parallels that of chicken malaria and further highlights the efficacy of these kinds of surface modified nanosilica. It is tempting however to speculate that this chicken and silkworm model for drug development might be used successfully for controlling other diseases of insect, animal and human diseases. In particular, controlling human malarial parasite growth both *in vitro* and *in vivo* might be possible[16-21]. All this may usher in a new era of `metabolomic modulator' based drug development using nanotechnology.

**METHODS**

All the Animal experiments were done in the animal facility of the West Bengal University of Animal and Fishery Sciences, West Bengal. 30 day old age matched Van Cobb breed broilers weighing approximately 250-300 gms from the M/S Gita Hatcheries (Barasat, West Bengal, India) were used throughout the study. The chickens were kept individually in standard cage with mosquito net wrapped around for preventing mosquito driven infection. The cages were kept in a well ventilated room maintained at $27^0$C. The Chickens were provided with standard feed and drinking water manufactured by Gita Hatcheries Ltd. and used in the hatchery regularly. Three birds per treatment and age matched controls (n=3) were used throughout the study and all protocols were approved by the animal ethics committee. *Plasmodium gallinaceum* Indian isolate frozen stocks were purchased from the parasite bank of the National Institute of Malaria Research (NIMR), New Delhi. The frozen stocks of the malarial parasite were thawed to $37^0$C and mixed with freshly collected chicken blood and parasitemia were determined quickly by standard smear technique. Equal volume of chicken blood inoculums (0.2 ml) containing less than 0.01% parasitemia were used for artificially infecting chickens via i. p. and i. v. injection. The parasitemia were observed by using blood smear examinations. At day 5, when the parasitemia were found to be very high, nanosilica dispersed in absolute alcohol were fed to the chicken orally (as drops and at the rate of 0.025mg/ml of nanosilica) in two doses of 0.5 ml each in the morning (8 am) and evening (8 pm). Higher and lower doses were also tried during the course of the treatment, but 0.025 mg /ml of nanosilica were found to be physiologically relevant as described in the results section. Data for the other doses have not been shown in this paper. Functionalized nanoporous silica were prepared from natural sources in collaboration with Fossilshield Inc., Germany, which are capable of physically absorbing chicken serum derived lipids (Fig. 1; panel V)[17-19].

Standard clinical, post mortem and pathological studies were done on liver, spleen, heart, brain and kidney of normal, infected and nanosilica treated infected chicken. These studies include changes in the clinical parameters, the gross morphological changes in the organs, histopathological studies and post-mortem changes in the aforementioned organs vis a vis control. Changes in the concentration of different isoenzymes, viz., peroxidase, esterase, lactate de-hydrogenase (LDH) and Glucose-6-phosphate-dehydrogenase. Changes in the metabolite levels like aspartate aminotransferase, glutamate dehydrogenase, γ-glutamyltransferase and creatinine were also studied. All the protocols for measuring isoenzymes and metabolites are described elsewhere in detail[20]. Serum from the control (day 1 and 7) as well as untreated infected birds (day 1 and day 7; after 8 days of parasite infection, the birds were treated with drugs) and infected chickens treated with different concentrations of nanosilica were collected and analyzed for their different cholesterol components (total Cholesterol, HDL, LDL, VLDL and serum triglycerides) with the professional help of M/S Pathowind Pathological laboratories Inc., India (Kolkata, West Bengal, India). Experiments were repeated at least three times in each case with at least three birds per treatment. Statistical analyses were performed using IRRISTAT software and data were potted following manufacturer's protocol.

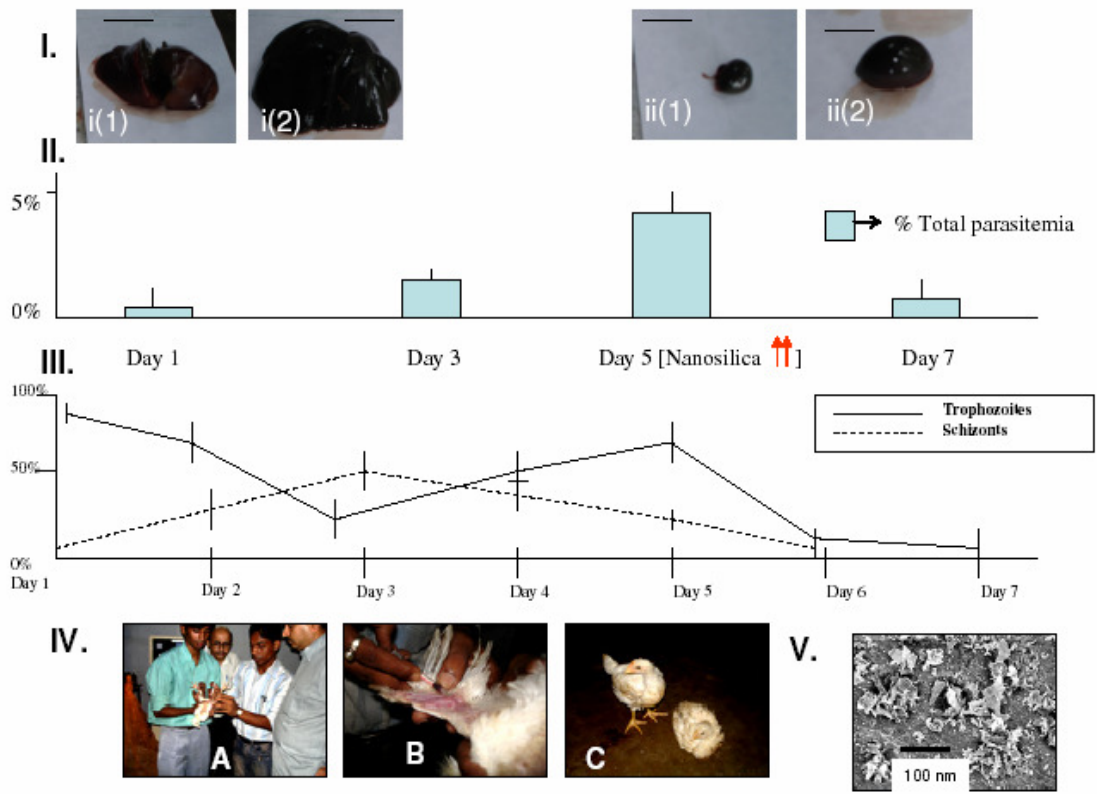

Fig. 1

Fig. 1 legend.

30 day old Vancobb breed broiler chickens were artificially infected with *P. gallinaceum* via i. p. (panel IV- A) and i. v. injection (panel IV-B). 7 days post infection, the parasitemia were found to be very high and the infected chickens showed typical clinical symptoms like posterior paralysis due to ischemia in brain (Panel IV-C). Panel II showed the hepato-splenomegaly in the infected chickens compared to control. Panel II shows the changes in the total parasitemia levels before and after the nanosilica treatment. Nanosilicas were administered 5 days post infection and within two days the treatment reduced the major parasitic load in chickens. Panel III shows the changes the levels of the two intra-erthrocytic stages of the parasite before and after the nanosilica treatment. Panel V shows the morphological characteristics of the nanosilica administered at the ultrastructural level. Details of these nano-structured silica is described elsewhere (12-14)

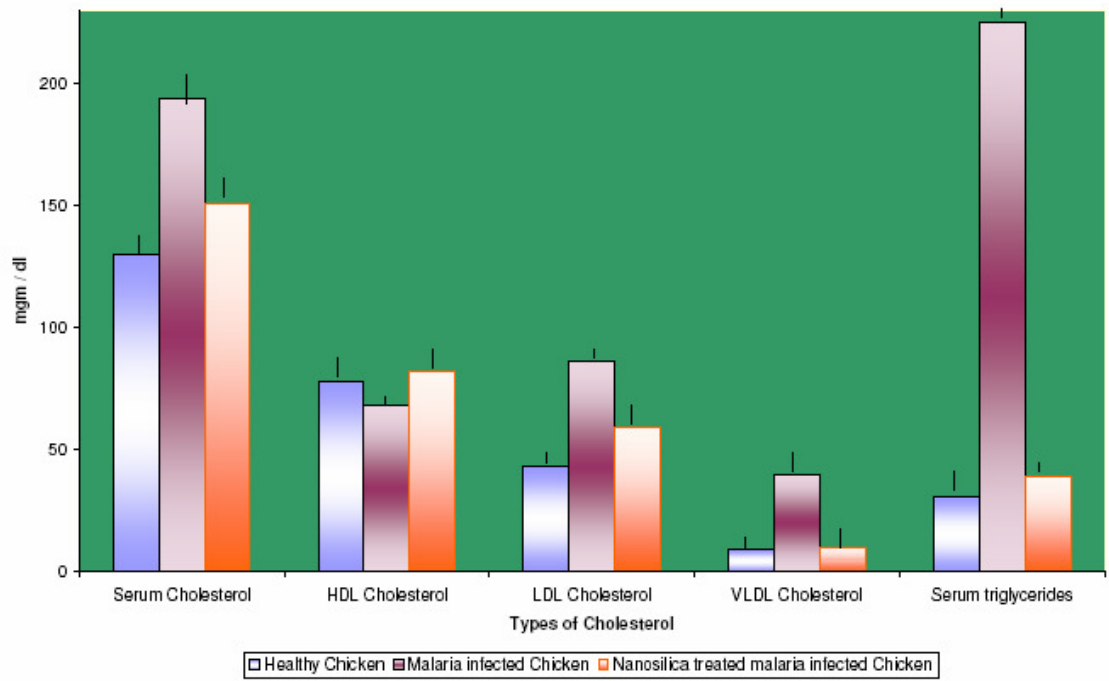

**Fig. 2**

Figure 2 legend.

Serum derived cholesterol and its various components were measured as described in the normal (uninfected), malaria infected and nanosilica drug treated malaria infected chickens. The mean total cholesterol level were found to be higher (194 ± 3.12 mgm / dl) compared to control (130 ± 4.46 mgm / dl). When the malaria infected chickens were treated with nanosilica 7 days post infection, the levels of the total cholesterol were found to be 151 ± 1.21 mgm /dl. Nanosilica treatment at the dose of 0.025 mg /ml causes marked reduction serum VLDL (Normal: 9 ± 1.12; Malaria infected:40 ± 2.13; Nanosilica treated: 10 ± 1.72) and triglycerides (Normal: 31 ± 2.47; Malaria infected:225 ± 3.07; Nanosilica treated:39 ± 5.47) level affecting the growth of the malarial parasite *in vivo*.

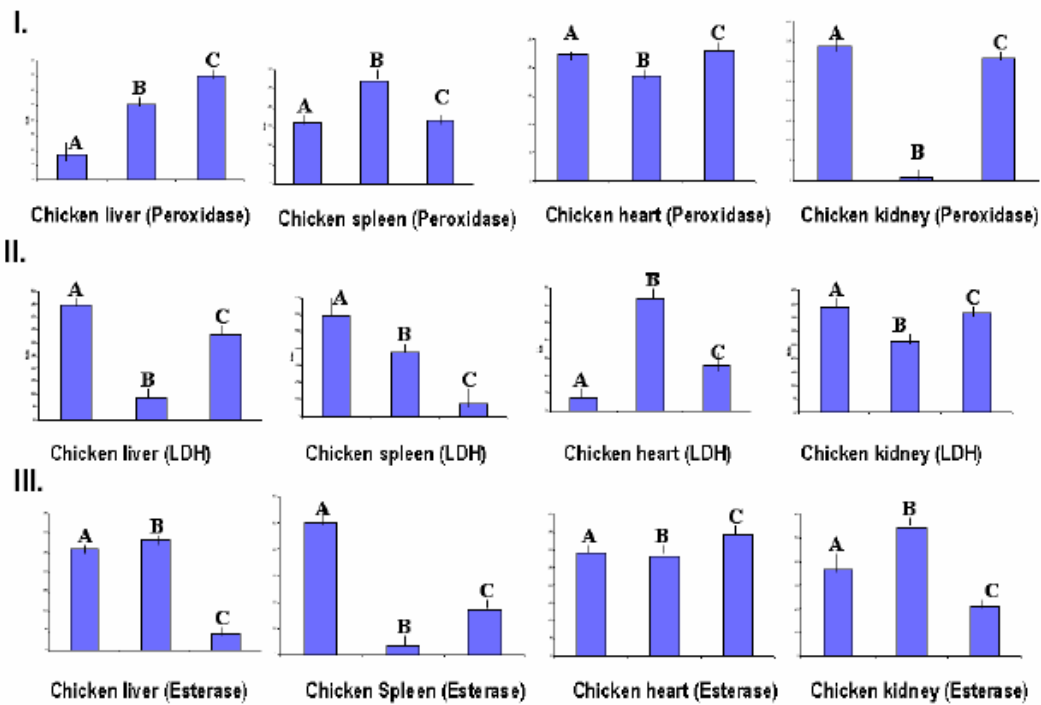

Fig. 3

Figure 3 legend.

Several isoenzymes levels were measured from the different tissues (spleen, liver, heart and kidney) of normal (panel A), malaria infected (panel B) and nanosilica treated (panel C) infected chickens (n=3) at day 7 post infection. The values were plotted with S.E.. Panel I shows that in the case of liver tissues, the concentration of the peroxidase enzyme increases significantly after administration of the nanosilica over control and malaria infected chickens. In case of spleen, The level of peroxidase increases with the malaria infection and nanosilica treatment reduces the level close to normal. In case of the heart tissue, the application does not cause significant changes in peroxidase level. In case of kidney tissues, the malaria infection causes highly significant reduction of the peroxidase level, but application of nanosilica increases the level of peroxidase. Panel II and III show the effect of nanosilica on *in vivo* levels of LDH and Esterases. This shows there are random changes in the levels of the enzymes indicating that there is no possibility of the bio-toxicity of these nanosilica inside chicken body.